\documentstyle[11pt,aaspp4]{article}

\newcommand{\rg}{r_{\rm g}}

\begin{document}
\title{Spinning Black Holes in AGN}
\author{Y. Dabrowski}
\affil{Astrophysics Group, Cavendish Laboratory, Madingley Road, Cambridge CB3 0HE, UK}

\begin{abstract}
Recent X-ray spectroscopy made with ASCA have shown broad, skewed iron
line emission from Seyfert-1 galaxies. The large extent of the red
tail allows probing of the innermost regions of the black hole's
accretion disc. A model of line emission has been developed and very
strong evidence for the presence of a rapidly rotating Kerr black hole
has been established in the case of MCG--6-30-15.  Issues related to
the observed line equivalent width and the position/geometry of the
primary source are discussed. Both the continuum and the reflected
iron line are computed, in a consistent manner, for a source located
on the axis of rotation.
\end{abstract}

\section{Introduction}

The iron line in MCG--6-30-15 was found to be both broad and skew from
a 4.5 day long observation in 1994 (Tanaka et al. 1995).  Much of the
skewness is explained by gravitational redshift which gives clear
evidence for the effect of strong gravity.  In Section
\ref{section:fits} we present briefly a simple model of line emission
from the accretion disc of a black hole. As explained in more details
in Dabrowski et al. (1997), we find that a fit to the line profile
observed in MCG--6-30-15 at minimum emission (Iwasawa et al. 1996)
gives very strong evidence for a rapidly rotating Kerr black hole.  In
Section \ref{section:images} we present a few images of the disc, as a
function of the observer's inclination angle and the black hole spin
parameter.  In Section \ref{section:ew} we investigate the model
proposed by Martocchia \& Matt (1996) and we predict the observed line
equivalent width in the case where the primary hard X-ray source is
located on the axis of symmetry, above the accretion disc. Finally, we
make a few comments about the forthcoming XMM mission in Section
\ref{section:xmm}.

\section{First model, fits to the line profile of MCG--6-30-15}
\label{section:fits}
\begin{figure}[htbp]
  \epsscale{1.0}
  \begin{center}
  \plotfiddle{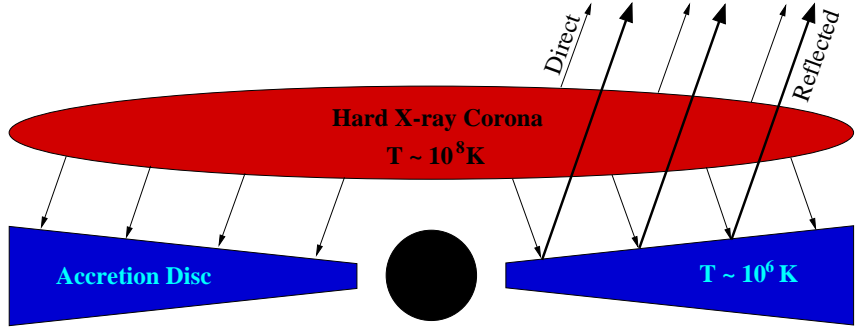}{75pt}{0}{80}{80}{-160}{-20} 
  \end{center}
  \caption{The central black hole is surrounded by an accretion disc
  of `cold' material. The observed power law spectrum (continuum)
  originates from the hot, hard X-ray corona assumed to be located
  above (and below) the disc. The illumination of the disc by hard
  X-ray photons is responsible for the observed reflected component,
  including strong fluorescent iron emission lines.}
\label{fig:geo1}
\end{figure}

The main parameters to which the line spectra are sensitive are the
inclination angle $i$ of the observer from the axis of symmetry, the
radial emissivity profile $\epsilon (r)$ of the disc and the angular
momentum $a/M$ of the hole. The black hole belongs to the Kerr family
and the metric used is in the Boyer-Lindquist form. The accretion disc
is assumed to be geometrically thin and axially symmetric around the
hole's axis of rotation. The fluorescence emission starts at the
radius of marginal stability $r_{ms}$, up to 15 gravitational radii
($15 \rg=15 GM/c^2$).  In order to concentrate on the influence of $i$
and $a/M$, the fluorescence emissivity $\epsilon (r)$ is assumed to
follow the law described in Page \& Thorne (1974), which closely
mimics a power law. However $\epsilon (r)$ vanishes at both $r_{ms}$
and infinity. The observer is located in the asymptotically flat
region (in practice at $10^4 \rg$).  The geometrical setup of the
system is described in Figure~(\ref{fig:geo1}).

Numerical integration of photon null geodesics are carried out
backwards in time, from the observer to the disc, taking into account
all the relativistic effects on both trajectory and redshift.  The
flux at the observer is computed using the phase-space occupation
number $I_{\nu}/\nu ^3$ which is invariant along the entire photon
trajectory.  In Dabrowski et al. (1997) we present various predicted
spectral lines over a range of $a/M$ and $i$. We found that the
overall width, and in particular the blue cut-off are most sensitive
to the inclination angle.  For sufficiently high inclination, the line
is double peaked, the red and blue peaks being due to the receding and
approaching components respectively.  In addition, the transverse
Doppler and gravitational effects tend to stretch the red wing of the
line spectrum.  This effect is boosted as the angular momentum
increases.  Indeed, high values of a/M allow the inner part of the
disc to lie closer to the hole, resulting in very strong gravitational
effects.

The evidence of strong gravity around a Kerr hole in MCG-6-30-15 has
already been discussed by e.g. Iwasawa et al. (1996).  Indeed, the
study of the variability of the fluorescent iron line profile during
the 4.5 day observation has revealed that it broadened still further
during a deep minimum in the light curve.  The red wing of the line
extends further to the low energies while the blue wing disappears.
This behaviour can be explained if much of the emission originates
from within $6 \rg$. Since a disc around a non-spinning Schwarzschild
black hole does not extend within this region, it is reasonable to
conclude that the black hole must be spinning.  We have carried out a
quantitative $\chi^2$ fit using the disc-line model presented above
and obtained joint constraints on both $a/M$ and $i$ (see Dabrowski et
al. 1997). The results give relatively good evidence for an
inclination angle of $\sim 25^o$ to $30^o$ and strongly favour an
extreme Kerr black hole ($a/M > 0.94$) rather than a Schwarzschild
black hole ($a/M = 0$).  The inclination $i\sim 30^o$ arises mainly
because of the observed cut-off for energies higher than 7~keV, while
the observed broad red wing necessitates the high value of $a/M$.
This result is the first observational evidence of rapidly rotating
black holes. From a theoretical point of view, their presence in
Active Galactic Nuclei (AGN) was formulated by e.g. K.S. Thorne in
1974.  The possibility that AGN harbour a rapidly rotating Kerr black
hole is most significant since the hole spin parameter is thought to
be the key parameter that would control the formation of relativistic
radio jets (e.g. Rees et al.1982).

\section{Accretion disc images}
\label{section:images}
\begin{figure}[h!]
  \epsscale{1.0}
  \begin{center}
  \plotfiddle{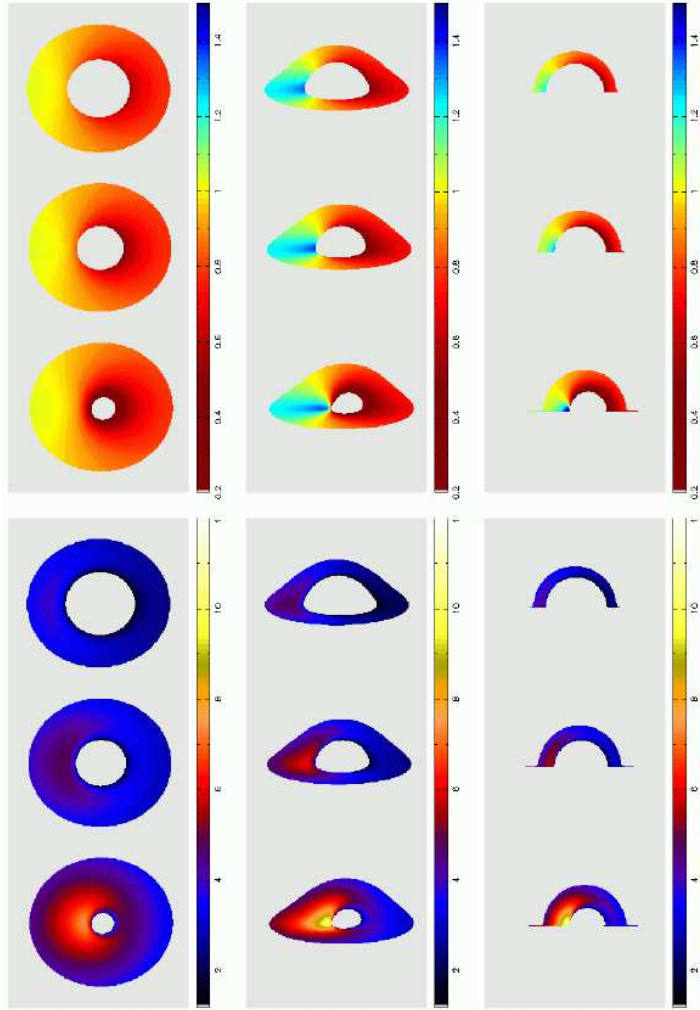}{330pt}{0}{80}{80}{-140}{-25}
  \end{center}
  \caption{ Images of accretion discs around Kerr black holes
  ($r_{in}=r_{ms}$, $r_{out}=15\rg$).  The first, second and third
  columns are for observers with inclination $i=30^o$, $75^o$ and
  $90^o$ respectively.  Each strip is composed of three images which
  correspond to $a/M=0$, $0.5$ and $0.998$, from top to bottom.  The
  colour-scale of the frequency strips (top row) indicates the
  variation of redshift ($\nu/\nu_0$). The colour-scale of the flux
  strips (bottom row) represents the logarithm of the observed flux.
  }
\label{fig:images}
\end{figure}

As mentioned above, given the inner and outer radii of emission
$r_{in}$ and $r_{out}$ and the emissivity law $\epsilon(r)$, disc
images depend upon the inclination $i$ and $a/M$.  As an illustration,
the images of Figure~\ref{fig:images} show the predicted frequency
contrast (top three strips) as well as the observed flux (bottom three
strips) of the line emission as seen by a distant observer.
It is clear, looking at the first column for example, that emission
can occur much closer to the central black hole in the case
$a/M=0.998$ than for a Schwarzschild black hole (a/M=0); furthermore,
this emission is highly redshifted.  Light bending due to strong
gravity is clearly visible on these images, particularly at large
inclination angles.  In the third column the disc is seen edge-on so
that one would expect its side only to be visible. What is actually
observed at $i=90^o$ is light from the top of the disc that has been
bent at an angle of $\pi/2$ towards the observer. Finally, looking at
the middle column for example, one notices that Kerr black holes with
$a/M=0.998$ allow for highly boosted emission from the blue shifted
approaching part of the disc.

\section{Improved model, problem of equivalent width}
\label{section:ew}
\begin{figure}[h!]
  \epsscale{1.0}
  \begin{center}
  \plotfiddle{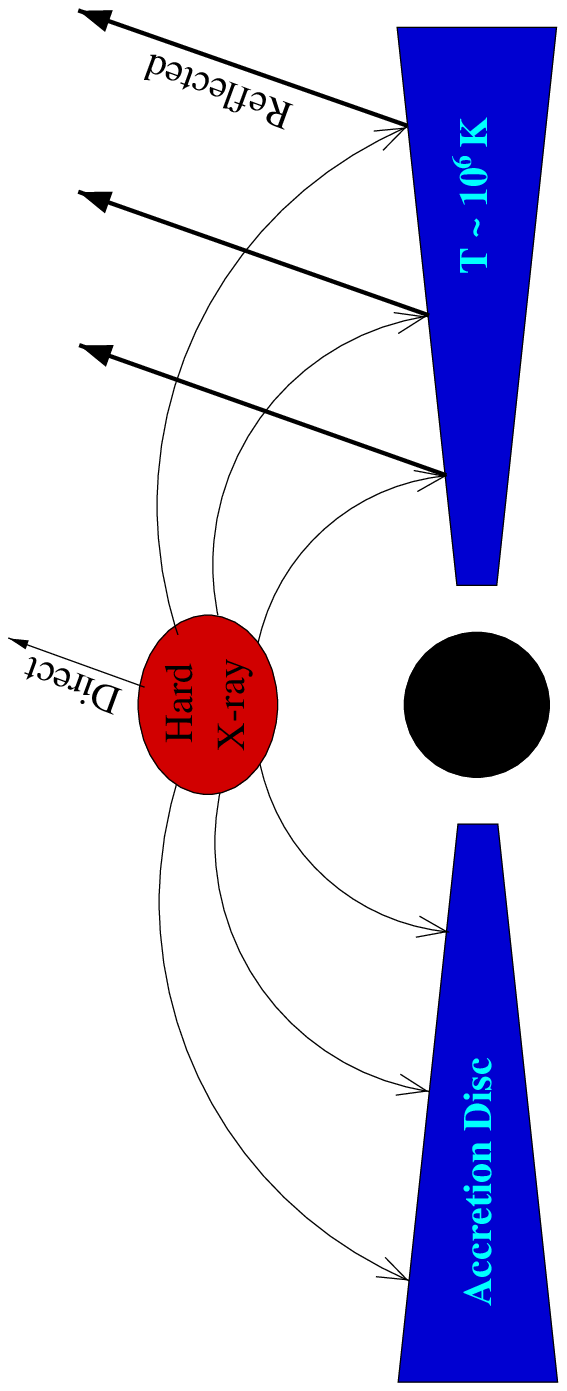}{80pt}{270}{80}{80}{-280}{250}
  \end{center}
  \caption{New geometry for the primary hard X-ray source. The
  accretion disc is here illuminated by a source of small extent,
  placed at height $h$ above the disc, on the axis of symmetry.}
\label{fig:geo2}
\end{figure}

\begin{figure}[h!]
  \epsscale{1.0}
  \begin{center}
  \plotfiddle{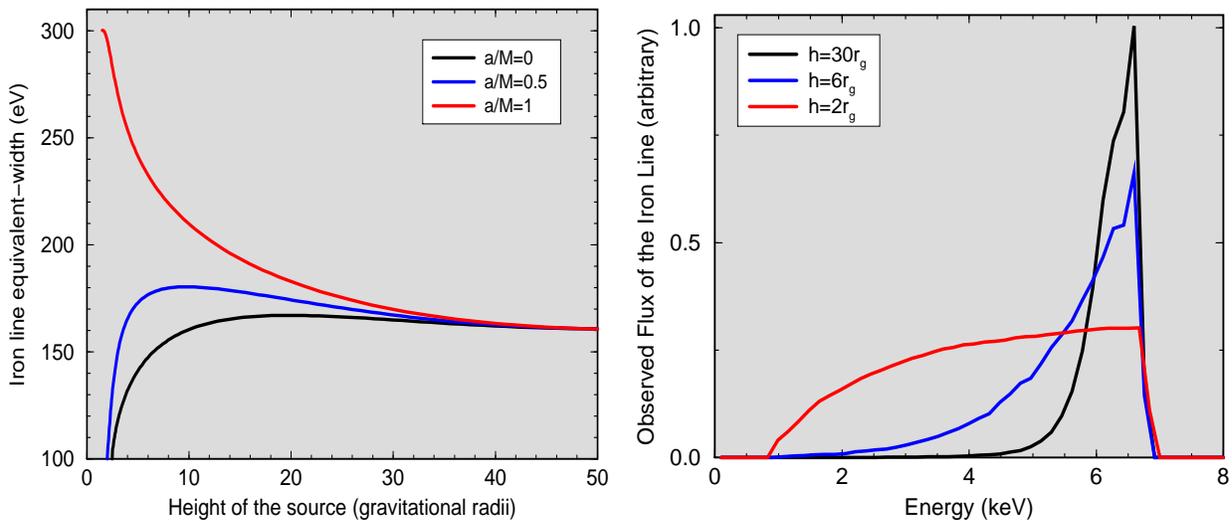}{150pt}{0}{80}{80}{-230}{-25}
  \end{center}
  \caption{ {\bf Left}: Measured iron line equivalent width as a
  function of the height $h$ of the primary source, for $a/M=0$, $0.5$
  and $1$. Neutral material with cosmic abundances is assumed (George
  \& Fabian, 1991). The Compton reflection continuum is not taken into
  account.  {\bf Right}: Observed iron line profiles for various
  source height.  $a/M=1$ and $i=30^o$ are assumed.  The flux
  magnitude is arbitrary, however the area of each line is
  proportional to its measured equivalent width.}
\label{fig:ew}
\end{figure}

For the above model, the line equivalent width is estimated to be
$\sim 100 - 200~{\rm eV}$ (see e.g. George \& Fabian 1991).  However,
as discussed in e.g. Nandra et al. (1997) for a sample of 18 Seyfert-1
galaxies, observations reveal equivalent widths of 300 - 600~eV, or
even as high as $\sim 1~{\rm keV}$ in the case of MCG--6-30-15.  In
order to address this issue, it is necessary to consider in more
detail the primary hard X-ray source, responsible for both the
continuum and the fluorescence emissions. In particular, as proposed
by Martocchia \& Matt (1996), the case where the source is located on
the rotation axis at a height $h$ above the accretion disc is
investigated here (see Figure~\ref{fig:geo2}).
If the source is close enough to the hole, most of the primary photons
are bent towards the disc, enhancing the reflected flux.  This effect
is negligible for the Schwarzschild case since a large amount of
radiation is lost in the event horizon.  However, for $a/M \sim 1$ the
enhancement should be significantly if the primary source is
sufficiently close to the hole ($h < 6 \rg$).  The fluorescence
process is taken into account following results given in George \&
Fabian (1991).  Monte Carlo simulations have been carried out for a
primary power law energy distribution $N(E)\propto E^{-1.7}$.  The
seed photons that hit the disc are reprocessed into iron fluorescent
photons ($6.4~{\rm keV}$) and both the primary and reflected photons
that escape to infinity are observed at $i=30^o$.  Consequently, both
the continuum and the reflected iron line components can be accounted
for in a consistent manner.  The fluorescent emission is assumed to
start at $r_{in}=r_{ms}$ up to $10^3 \rg$ and the observer is located
at $10^5\rg$. The predicted equivalent width $W_K$ is found to be
$\sim 160~{\rm eV}$ for a sufficiently distant source ($h > 30
\rg$), which is in agreement with e.g. George \& Fabian (1991) and
Matt et al. (1992).  As expected, in the case of primary sources
located closer to the black hole, $a/M$ plays a fundamental role (see
Figure~\ref{fig:ew}).
In the case of slowly rotating black holes, no significant equivalent
width enhancement is obtained.  For rapidly rotating holes, the strong
gravitational focusing of light rays combined with emission from the
very inner regions contribute to the enhancement of $W_K$.  However,
the highest equivalent width obtained is only 300~eV for the extreme
case where $a/M=1$ ($r_{ms}=1.23 \rg$) and $h=2 \rg$.

\section{The X-ray Multi-mirror Mission}
\label{section:xmm}

The XMM satellite is several times more sensitive than ASCA, for
example.  We therefore expect that this new X-ray mission will yield
significant returns in understanding AGN central engine and focusing
on to properties of supermassive black holes.  High resolution
spectroscopy will be carried out on many AGN.  This will allow more
precise study of the iron line profile from samples of AGN larger than
considered so far (e.g. Nandra et al. 1997, Bromley et al. 1998).
Most of the results concerning black holes characteristics are model
dependent and XMM will help in discriminating between different
scenarios such as, for example, the occultation model of Weaver \&
Yaqoob (1998), or the possibility of iron fluorescence from within the
minimum stable orbit $r_{ms}$ proposed by Reynolds \& Begelman (1997).
Finally recent work by Reynolds et al. (1998), on the temporal
response of the iron line to individual activation/flaring of X-ray
emitting region, shows how a mission such as XMM will help in
constraining the system geometry as well as the black hole mass and
angular momentum (see also Fabian et al., this proceedings).

\section{Acknowledgements}
The work presented here, and particularly in
Section~\ref{section:fits}, is in collaboration with A.C. Fabian,
K. Iwasawa, A.N. Lasenby and C.S. Reynolds.


\begin{thebibliography}{}
\bibitem[]{}
Bromley, B.C., Miller, W.A., Pariev, V.I. 1998, Nature, 391, 54
\bibitem[]{}
Dabrowski, Y., Fabian, A.C., Iwasawa, K., Lasenby, A.N., Reynolds, C.S. 1997, MNRAS, 288, 
L11
\bibitem[]{}
George, I.M., Fabian, A.C. 1991, MNRAS, 249, 352
\bibitem[]{}
Iwasawa, K. et al. 1996, MNRAS, 282, 1038
\bibitem[]{}
Martocchia, A., Matt, G. 1996, MNRAS, 282, L53
\bibitem[]{}
Matt, G., Perola, G.C., Piro, L., Stella, L. 1992, A\&A, 257, 63
\bibitem[]{}
Nandra, K., George, I.M., Mushotzky, R.F., Turner, T.J., Yaqoob, T. 1997, ApJ, 477, 602
\bibitem[]{}
Page, D.N., Thorne, K.S. 1974, ApJ, 499, 191
\bibitem[]{}
Rees, M.J., Begelman, M.C., Blandford, R.D., Phinney, E.S. 1982, Nature, 295, 17
\bibitem[]{}
Reynolds, C.S., Begelman, M.C. 1997, ApJ, 488, 109
\bibitem[]{}
Reynolds, C.S., Young, A.J., Begelman, M.C., Fabian, A.C. 1998, astro-ph/9806327
\bibitem[]{}
Tanaka, Y. et al. 1995, Nature, 375, 659
\bibitem[]{}
Thorne, K.S. 1974, ApJ, 191, 507
\bibitem[]{}
Weaver, K.A., Yaqoob, T. 1998, ApJ, 502, L139
\end{thebibliography}
\end{document}